# Impact Analysis of Utility-Scale Energy Storage on the ERCOT Grid in Reducing Renewable Generation Curtailments and Emissions

Cody Buehner, Sharaf K. Magableh, Oraib Dawaghreh, Caisheng Wang
*Department of Electrical and Computer Engineering*
*Wayne State University, Detroit, United States*
cody.buehner@wayne.edu, sharaf.magableh@wayne.edu, oraib.dawaghreh@wayne.edu, cwang@wayne.edu

*Abstract*—This paper explores the solutions for minimizing renewable energy (RE) curtailment in the Texas Electric Reliability Council of Texas (ERCOT) grid. By utilizing current and future planning data from ERCOT and the System Advisor Model from the National Renewable Energy Laboratory, we examine how future renewable energy (RE) initiatives, combined with utility-scale energy storage, can reduce $CO_2$ emissions while reshaping Texas's energy mix. The study projects the energy landscape from 2023 to 2033, considering the planned phase-out of fossil fuel plants and the integration of new wind/solar projects. By comparing emissions under different load scenarios, with and without storage, we demonstrate storage's role in optimizing RE utilization. The findings of this paper provide actionable guidance for energy stakeholders, underscoring the need to expand wind and solar projects with strategic storage solutions to maximize Texas's RE capacity and substantially reduce $CO_2$ emissions.

## I. Introduction

Texas is a major player in both consuming and producing fossil fuels [1]. It accounts for nearly 12% of the total energy used in the United States. The state's hot and humid climate drives significant demand for air conditioning during the summer months. Historically, low energy prices, abundant crude oil and natural gas reserves, a robust energy infrastructure, and a growing economy have made Texas a prime location for energy-intensive industries. In February 2021, Texas's winter storm highlighted the challenges and potential solutions for RE curtailment, particularly emphasizing the role of energy storage systems (ESS) in maintaining grid stability during extreme weather [2]. ERCOT's limited interconnectivity with the national grid restricted its ability to import electricity, placing added pressure on its internal generation resources [2]. As ERCOT advances toward a cleaner energy mix dominated by wind and solar, energy storage systems (ESSs) emerge as a key asset to reduce renewable curtailments [3]. The Annual Energy Outlook 2023 suggests that Texas could effectively address intraregional transmission bottlenecks that contribute to renewable curtailments with continued investments in transmission infrastructure and expanded storage capacity. Moreover, strategic ESS deployment will enable ERCOT to minimize curtailments while reducing reliance on gas peakers [4], contributing to a cleaner, more resilient grid. Given the expected increase in severe weather due to climate change, this approach provides a foundation for meeting resilience goals and aligns with ongoing policies prioritizing RE expansion [5].

To clarify the research landscape for this study, the following literature review examines relevant studies and findings. In [6], the researchers investigate how transmission line limitations affect RE curtailments in Texas as a case study by the U.S. Energy Information Administration. Using the UPLAN model to simulate energy procurement, pricing, and constraints, the study was calibrated with 2022 ERCOT generation data and projected out to 2035. Importantly, the study estimated additional renewable generators by analyzing the Generator Interconnection Status (GIS) Report from ERCOT and determining which wind and solar projects are most likely to be completed in the interconnection process. In [7], the authors study the timescales of ESS needed to reduce RE curtailment including various mixes of wind and solar renewable energy resources (RERs) to account for up to 55% of electricity generation in ERCOT. The study observed a significant increase in energy curtailments when RE penetration in ERCOT's grid approached 55%. The authors in [8] explored the role of energy storage technology in supporting Texas's transition to increased RER, focusing on behind-the-meter commercial microgrids. Their investigation uses ERCOT's 2022 data to assess renewable penetration across various Texas regions. However, a limitation in their work is the narrow scope of analyzing ESS solely from a commercial user's perspective, which may overlook larger, grid-wide challenges related to energy reliability, curtailment, and emissions. Additionally, the study's evaluation framework lacks a comprehensive view of long-term renewable integration beyond current load conditions, potentially limiting its applicability to broader energy policy decisions. Recent studies have demonstrated how integrating network constraints can improve the feasibility of emission-based dispatch strategies. For instance, authors in [9] proposed a two-stage robust power dispatch model that incorporates carbon trading, ensuring that dispatch decisions remain technically viable while minimizing operating costs, risk costs, and carbon emissions. This model effectively addresses uncertainties in RE output and enhances RE utilization, demonstrating the importance of robust, network-aware dispatch strategies. Additionally, in [10], the researchers conducted a comprehensive review of ESSs in decarbonizing power systems, highlighting their critical role in mitigating transmission congestion, reducing renewable curtailment, and supporting emission-reduction goals. Their findings emphasize the need for dispatch models to account for the technical characteristics of ESSs and local grid constraints to achieve cost-effective and sustainable power system operations. These studies underscore the importance of considering grid constraints and storage integration when designing dispatch models for renewable-dominated power systems. In contrast, our work builds on the ERCOT data but expands the scope to a nationwide, utility-scale perspective. By analyzing both current and projected RE initiatives from 2023 to 2033, we investigate how future grid-scale storage solutions can minimize renewable curtailment and emissions across Texas. Fig. 1 illustrates an overview of the ERCOT grid and its system configuration.

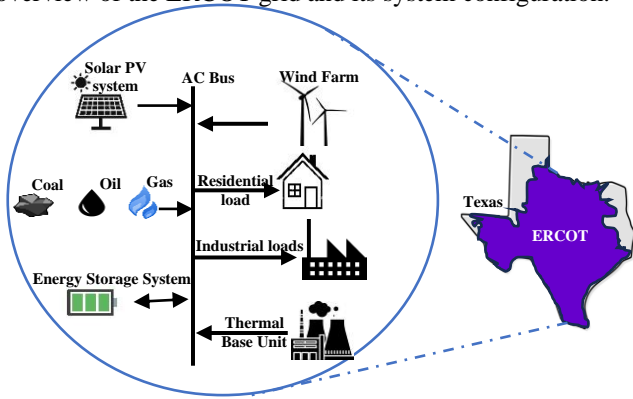

Fig. 1. Schematic representation of the ERCOT grid and configuration system.

This paper investigates the effect that grid energy storage has on $CO_2$ emissions, using ERCOT as a case study. Our study uses a similar approach as [6] to predict the overall future ERCOT wind and solar capacity, but focuses on curtailments caused by RE over production rather than transmission constraints, and crucially, how overall carbon intensity of the electrical generation is impacted by these curtailments. In order to maximize RE utilization, a unique dispatch model is proposed which ranks various generation sources based on CO2 emissions and prioritizes low carbon sources 1$^{st}$ to meet load. The methodology stands out by integrating ERCOT's planning data with the System Advisor Model (SAM) from NREL to simulate the hourly output from each future wind and solar project within the ERCOT system. By using both current and projected load and generation data, strategically dispatching ESS can help minimize renewable curtailments while enabling greater adoption of wind and solar energy. Unlike conventional storage approaches, which are typically designed for isolated load smoothing or peak-shaving applications, our analysis underscores the strategic role of storage in actively displacing fossil fuel-based generation by storing excess RE during low-demand periods and supplying it during peak demand. This approach will enhance renewable utilization and directly contribute to reducing grid-level $CO_2$ emissions by minimizing the reliance on fossil fuels during high-demand periods. Our findings emphasize the specific value of storage in supporting a projected increase in renewable penetration up to 55% by 2033, illustrating its critical function in aligning future energy generation with emissions reduction goals for a more sustainable Texas energy landscape.

## II. DESIGN AND MODELING APPROACH

The mathematical modeling framework used in this study is based on an optimization-driven dispatch model that prioritizes generation resources based on carbon intensity while integrating ESS to minimize renewable energy curtailment. The dispatch model follows a sequential approach, starting with the calculation of total renewable and non-renewable generation available for the given hour, then dispatching generation based on type with priority given to low carbon sources. The objective function minimizes total system emissions while ensuring that generation meets the hourly demand. Wind and solar curtailment are computed as the difference between available renewable generation and the dispatched portion. To account for the role of energy storage, a secondary optimization step reallocates curtailed energy into storage and discharges it when demand exceeds renewable supply. The emissions reduction is quantified using carbon intensity factors associated with different generation sources. The mathematical formulation of this process includes constraints on energy balance, storage capacity, and dispatch priorities, ensuring a realistic representation of grid operations under different scenarios. More information about the model formulation is referred to [11].

In this section, the methodology used to evaluate the impact of grid ESS on RE curtailment and $CO_2$ emissions within the ERCOT system is explained. By comparing a baseline scenario with several simulated future scenarios, we can assess how integrating ESS with RERs alters emission outcomes. This process is critical in understanding how different energy dispatch strategies, especially those involving renewable curtailment and energy storage, influence grid emissions and help optimize RE utilization.

*A. Baseline Scenario*

$CO_2$ emissions were determined for the baseline case by multiplying the values in Table I with the hourly generation from ERCOT's Fuel Mix Report for 2023. The hourly generation data reflects the actual output of various generation sources, including renewable, fossil fuel, and nuclear power, thereby allowing for an accurate representation of the carbon emissions attributable to ERCOT's energy production. By integrating the carbon intensity factors-expressed in kg of $CO_2$ emitted per kWh (kg$CO_2$/kWh)-with the generation volumes, a comprehensive estimate of total $CO_2$ emissions for the baseline scenario was derived. This baseline serves as a critical reference point for evaluating the effectiveness of subsequent simulations and the overall impact of ESS on reducing emissions in ERCOT.

TABLE I. CARBON INTENSITY OF VARIOUS GENERATION RESOURCES (KG$CO_2$/KWH) [12].

| Fuel | $CO_2$ (kg)/kWh |
|---|---|
| Biomass | 0.28 |
| Coal | 0.92 |
| Natural Gas Turbine | 0.55 |
| Natural Gas Combined Cycle | 0.44 |
| Hydroelectric | 0.024 |
| Nuclear | 0.012 |
| Photovoltaic | 0.026 |
| Wind | 0.011 |

*B. Simulated Scenarios*

The following process steps were used to calculate $CO_2$ emissions for all simulated scenarios. This process is also illustrated in Fig. 2.

1. Determine hourly capacity for non-wind and non-solar resources: For all simulations, it was assumed that the capacity of all resources other than wind and solar remained the same as in the baseline case.
2. Determine total hourly wind and solar generation: The hourly generation for future wind and solar projects was

simulated using the System Advisor Model (SAM) from NREL, with county-level average hourly weather data. This was added to the existing baseline wind and solar generation to provide the total available wind and solar generation for each hour of the year.
3. Dispatch available generation to meet load: All available generation resources were ranked by their carbon intensity (kg $CO_2$/kWh) from lowest to highest, based on Table I. The model dispatched the lowest-carbon resources first, continuing down the list until sufficient generation was allocated to meet the hourly load. Storage was not considered at this point.
4. Calculate wind and solar curtailment: In hours of low load, the combined generation of low-carbon (non-fossil fuel) resources sometimes exceeds the demand. Any excess wind and solar production was treated as curtailed since storage was not included in this stage.
5. Determine hourly $CO_2$/kWh without storage: Based on the dispatch results from step 3, $CO_2$ emissions per kWh were calculated using the carbon intensity values from Table I.
6. Re-run dispatch model with storage: The dispatch model was then run again, this time incorporating storage. The curtailment calculated in step 4 was used to determine the storage capacity available for dispatch in each hour. This capacity was treated as a resource and dispatched before relying on more carbon-intensive options such as natural gas. Unlike other resources, however, dispatching storage reduced its available capacity.
7. Determine hourly $CO_2$/kWh with storage: Once the generation profile, including storage, was known, the $CO_2$ emissions were recalculated to assess the impact of storage.

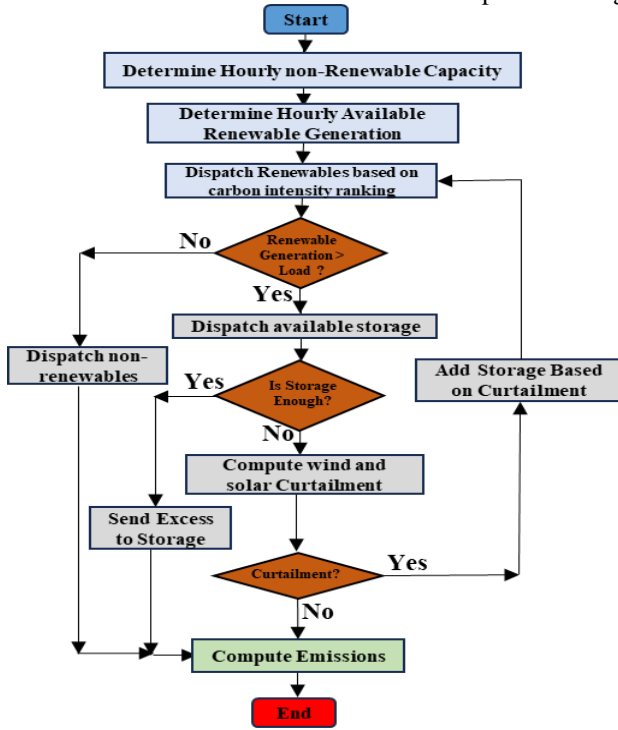

Fig. 2. Simplified flowchart of the system modeling process.

To summarize the previous process, this study evaluates the impact of grid-scale storage on emissions reduction and renewable energy utilization. By designing a set of scenarios representing different load profiles and storage configurations, the dispatch model prioritizes non-wind and non-solar resources before including available wind and solar generation. The model then allocates energy based on carbon intensity rankings, ensuring the lowest-carbon resources are dispatched first. Subsequently, to assess the effect of storage, we first calculate curtailment without storage and then re-run the model with storage capacity derived from curtailment levels. This comparative approach allows us to quantify the role of storage in mitigating curtailment and optimizing emissions reductions.

### III. SYSTEM'S DATA AND PROCESSING

Emissions in any power system are primarily determined by two factors: the load (or demand) on the system and the energy generation mix used to meet that demand [13]. In this paper, two hourly load profiles were chosen: actual total system load from ERCOT for 2023 as illustrated in Fig. 3 and 2033 projected load also from ERCOT [14].

The baseline generation was derived from ERCOT's 2023 Fuel Mix Report [15], which provides 15-minute interval data. For each hour, four 15-minute values were averaged to create an hourly generation profile for the entire year. Future capacity for all generation sources, excluding wind and solar, was based on ERCOT's Capacity and Reserves Report [16].

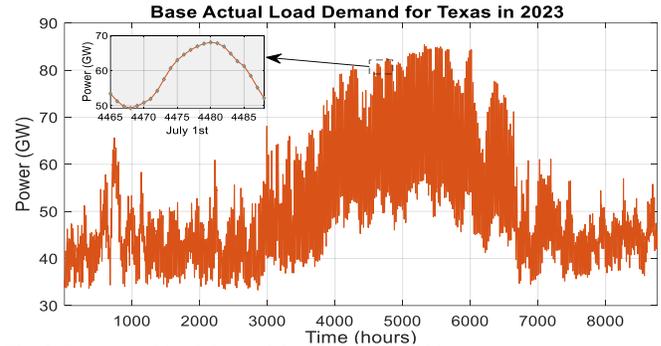

Fig. 3. Base actual load demand for ERCOT in 2023.

Figs. 4 and 5 present a comparison of baseline and forecasted solar and wind power generation profiles, providing a clear visualization of the increasing importance of RERs. Future wind and solar capacities were determined by choosing future wind and solar projects likely to come online in the near future from ERCOT's GIS report for June 2024 [17] and [18] and assuming this additional capacity is available for dispatch by 2033.

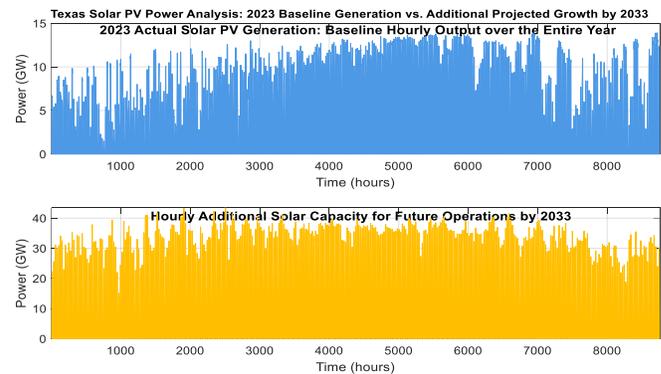

Fig. 4. Analysis of Texas solar PV power: 2023 baseline generation vs. additional projected growth by 2033.

As noted in [6], a 3-step approval process is required for all large (> 1 MW) generators wishing to connect to the ERCOT grid, so the GIS report provides insight into probable additional generator capacity in the future. Projects further along in the process are more likely to come to fruition, but the choice of which projects to include greatly impacts the predicted future capacity and thus future emissions. For this analysis, wind and solar projects from the GIS report were included if they met the following criteria: Security screening complete, interconnect study complete, or security screening complete, full interconnect survey in progress, and interconnection agreement complete. Applying these criteria resulted in an estimated additional future capacity of 59 GW of solar and 11 GW of wind.

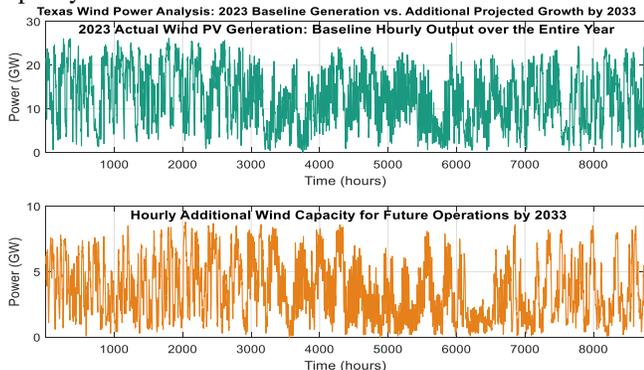

Fig. 5. Analysis of Texas wind power: 2023 baseline generation vs. additional projected growth by 2033.

Accurate predictions of these additional wind and solar projects require localized weather data, which was provided at the county level based on the information in the GIS report. Typical Meteorological Year data from the National Solar Radiation Database was used for all solar projects and 2014 wind data (the latest available) from the NREL Wind Toolkit was used for all wind projects. The assumed $CO_2$ emissions for each source are based on the life cycle emissions per kWh, as presented previously in Table I. Table II presents the comparative statistics (minimum, mean, and maximum) for base load, solar PV, and wind power generation in 2023 and 2033, highlighting the variations in RE output over time.

TABLE II. COMPARATIVE STATISTICS OF POWER GENERATION (MIN, MEAN, MAX) IN (GW) FOR BASE LOAD, PV, AND WIND FOR 2023 AND 2033.

| Power Type (GW) | | Minimum | Mean | Maximum |
| --- | --- | --- | --- | --- |
| Base Load 2023 | | 33.3 | 50.7491 | 85.46 |
| PV | 2023 | 0 | 3.6992 | 13.930 |
| | 2033 | 0 | 10.2923 | 43.4 |
| Wind | 2023 | 0.01 | 12.3282 | 26.2 |
| | 2033 | 0.01 | 3.5954 | 8.84 |

## IV. ANALYSIS AND IMPLICATIONS

In this section, we present the results derived from analyzing wind power generation data and explore the impacts of integrating ESS solutions into the ERCOT grid. The projected load demand from ERCOT is illustrated in Fig. 6, followed by an examination of generation mix variations, emissions trends, and storage utilization scenarios.

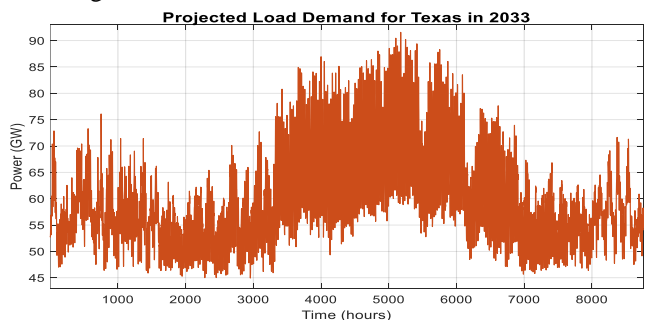

Fig. 6. Projected load demand in ERCOT for 2033.

The data is analyzed for both current and projected future scenarios, highlighting the contributions of RERs to total generation and the role of storage in curtailment management and emission reductions.

### A. Generation Mix

The total yearly generation for the baseline scenario, 2023 Load No Storage, and 2023 Load with Storage is consistent at 445 TWh, as all these scenarios utilize the same load profile. In contrast, the total generation for the two 2033 scenarios is projected to be higher at 530 TWh, reflecting the anticipated increase in load. Approximately 14% of the electricity generated in the baseline case comes from coal. However, the increase in wind and solar generation, especially solar, in all other scenarios is sufficient to eliminate coal from the generation mix, as the dispatch model prioritizes resources based on their carbon intensity. Furthermore, solar and wind accounted for roughly 30% of the total electricity generation in the 2023 baseline scenario, which does not include storage, as shown in Fig. 7. When factoring in additional wind and solar projects, RE contributions rise to between 48% and 55%, depending on the load profile. Table III also contrasts scenarios with and without storage, detailing total storage capacity (GWh) and the percentage of hours storage was utilized.

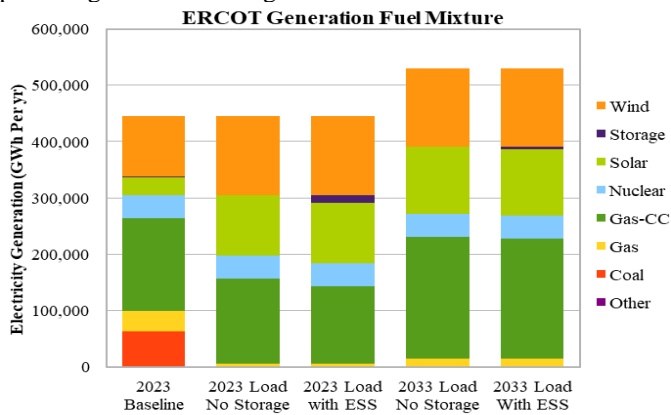

Fig. 7. ERCOT generation fuel mixture.

The scenario-based approach in our study provides valuable insights into how storage interacts with varying load conditions. As seen in Table III, the addition of storage increases renewable energy penetration by reducing curtailment and displacing fossil fuel-based generation. However, the extent of storage utilization varies across scenarios due to differences in curtailment hours. In the 2023 load profile, storage operates in 14.5% of all hours, whereas in the 2033 scenario, increased load demand naturally reduces curtailment, leading to a lower storage utilization rate of 4.33%. These findings emphasize that while storage plays a crucial role in integrating renewables, its effectiveness is influenced by the overall generation mix and demand patterns. Future expansions in both renewable capacity and storage must consider these dynamics to maximize emissions reductions and grid reliability.

TABLE III. WIND, SOLAR, AND STORAGE USAGE ACROSS DIFFERENT SCENARIOS.

| Scenario | Renewables % over Total Generation | Storage (GWh) | % of Hours Storage Used |
| --- | --- | --- | --- |
| 2023 Baseline | 31.49 % | N/A | N/A |
| 2023 Load with Storage | 55.44 % | 34.28 | 14.47 % |
| 2033 Load with Storage | 48.72 % | 33.67 | 4.33 % |

## B. Emissions

Eliminating coal from the generation mix is the largest contributor to the reduced carbon intensity from the baseline observed in Fig. 8. The integration of storage further lowers overall emissions by an additional 4.6% when utilizing the 2023 load profile. In both load scenarios, storage contributions effectively offset generation that would otherwise come from natural gas turbines or natural gas combined cycle generators. Although emissions for the 2033 load profiles are higher compared to 2023, they remain below the baseline levels. The modest curtailment in the 2023 profile allows wind and solar generation to operate near full capacity for the majority of hours. As the load increases in 2033, it primarily absorbs this excess wind and solar capacity, with any remaining demand being met through natural gas generation. This shift leads to an overall increase in emissions relative to 2023, highlighting the complexities of integrating renewables while managing load demands.

## C. Storage Capacity

Table III shows the amount of storage capacity needed to remove all the simulated curtailments. Although the total storage capacities needed for 2023 and 2033 are similar, the percentage of hours of utilization is significantly lower in the 2033 load profile. This indicates that the peak curtailment is similar in both load profiles, but curtailment occurs much less frequently.

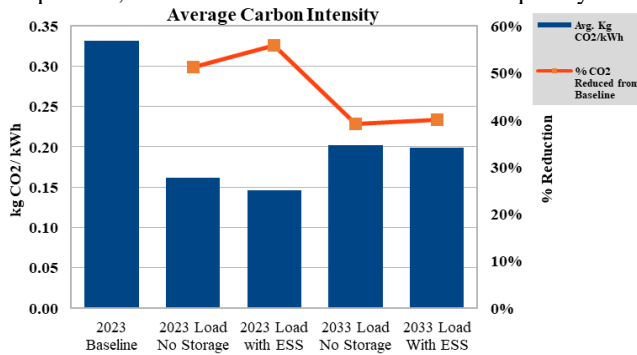

Fig. 8. Average carbon intensity of electricity generation.

## V. CONCLUSIONS

This study demonstrates that projected wind and solar growth could reduce ERCOT's $CO_2$ emissions by up to 51% by 2033, driven by 59 GW of solar and 11 GW of wind capacity replacing coal. Coupled with 30 GWh of storage, carbon intensity could decrease an additional 4.5%. Wind and Solar's share of generation could rise from 30% to 55%, eliminating coal and reducing reliance on natural gas. Storage, though representing only 3% of total generation capacity in 2023, mitigates curtailment and offsets gas-powered generation, becoming increasingly vital as renewable capacity expands. Our analysis focused on projects nearing interconnection approval; including early-stage solar projects could double capacity, altering curtailment dynamics and enhancing storage's role. As renewable generation grows, curtailment risks rise, but strategic storage deployment can maximize RE utilization and minimize fossil fuel dependence. Future research should explore broader project pipelines and varying curtailment levels to refine storage's potential in optimizing emissions reductions.

These findings underscore the urgency of pairing renewable expansion with targeted storage investments. Policymakers and stakeholders must prioritize integrated planning to avoid curtailment inefficiencies and accelerate emissions reductions. By aligning infrastructure development with renewable growth, Texas can fully leverage its RE potential, achieving both grid reliability and climate goals.

## ACKNOWLEDGMENT

This work was supported in part by the National Science Foundation of USA under Grant ECCS-2146615 and partially supported by the Department of Energy, Solar Energy Technologies Office Renewables Advancing Community Energy Resilience program under Award Number DE-EE0010413. Any opinions, findings, conclusions, or recommendations expressed in this material are those of the authors and do not necessarily reflect the views of the Department of Energy.